\documentclass[useAMS,usenatbib,psfig]{mn2e}

\usepackage{epsfig}
\usepackage{times}
\usepackage{graphicx}
\usepackage{color}
\usepackage{lscape}

\newcommand{\bd}{\begin{displaymath}}
\newcommand{\ed}{\end{displaymath}}
\newcommand{\be}{\begin{equation}}
\newcommand{\ee}{\end{equation}}

\newcommand{\msun}{{\rm M}_{\sun}}

\title{The black hole mass, Eddington ratio, and $M_{\rm bh}-\sigma_{[\rm O\ III]}$ relation in young radio galaxies}

\author[Q. W. Wu]
{ Qingwen Wu
\thanks{E-mail: qwwu@shao.ac.cn}\\
International Center for Astrophysics, Korea Astronomy and Space
Science Institute, 61-1, Hwaam-Dong, Yuseong-Gu, Daejeon 305-348,
Republic of Korea}

\date{Accepted 2009 May 20. Received 2009 May 20; in original form 2009 March 9}

\pagerange{\pageref{firstpage}--\pageref{lastpage}}
\pubyear{}

\begin{document}

\maketitle
\label{firstpage}

\begin{abstract}

The masses of the central supermassive black holes and the Eddington
ratios for a sample of 65 young radio galaxies [27 gigahertz-peaked
spectrum (GPS) and 38 compact steep-spectrum (CSS) sources] are
estimated by various methods. We find that the average BH mass of
these young radio galaxies is $<$$\log M_{\rm bh}$$>$$\simeq8.3$,
which is less than that of radio loud QSOs and low redshift radio
galaxies ($<$$\log M_{\rm bh}$$>$$\simeq9.0$). The CSS/GPS sources
have relatively high Eddington ratios with average ratio $<$$\log
L_{\rm bol}/L_{\rm Edd}$$>=-0.56$, which are similar to those of
narrow line Seyfert 1 galaxies (NLS1s). It suggests that the CSS/GPS
sources may not only be in the early stage of radio activities, but
also in the early stage of their accretion activities.

We find that the young radio galaxies as a class deviate
systematically from the $M_{\rm bh}-\sigma_{*}$ relation defined by
nearby inactive galaxies, when using $\sigma_{\rm [O\ III]}$ as a
surrogate for the stellar velocity dispersion $\sigma_{*}$. There is
no significant correlation between the deviation of the [O III]
emission line width, $\Delta\sigma\equiv\sigma_{\rm [O\
III]}-\sigma_{\rm [pred]}$, and the jet/accretion power, where
$\sigma_{\rm [pred]}$ are calculated from the Tremaine et al.
relation using the estimated BH masses. However, we find that the
deviation $\Delta\sigma$ in young radio galaxies is well correlated
with the Eddington ratio, and this correlation is found to be
similar to that of radio quiet AGN (QSOs/NLS1s) where the radio jet
is absent or weak. We suggest that the accretion activities may
still play an important role in shaping the kinematics of [O III]
narrow line in these young radio galaxies.

\end{abstract}


\section{Introduction}
Gigahertz-peaked spectrum (GPS) and compact steep-spectrum (CSS)
radio sources constitute a large fraction ($\sim40\%$) of the
powerful ($\log P_{1.4}\geq10^{25}\rm W\ Hz^{-1}$) radio source
population. Their radio spectra are simple and convex with peaks
close to 1 GHz and 100 MHz for GPS and CSS sources respectively.
Morphologically, the GPS sources are generally less than 1 kpc in
projected linear size and CSS sources are less than 20 kpc (see
O'Dea 1998 for a review). Two main theories have been proposed to
explain the CSS/GPS phenomenon: the \emph{youth scenario} (Fanti et
al. 1995) and the \emph{frustration scenario} (van Breugel 1984).
The age of CSS/GPS sources has been estimated using both the lobe
proper motion (e.g., Owsianik \& Conway 1998) and the radiative ages
(Murgia et al. 1999). The ages derived are less than a few thousand
years, therefore strongly supporting the youth scenario in the
interpretation of their compactness. In fact, it has been suggested
that compact CSS/GPS sources perhaps eventually evolve into
extended, edge-brightened Fanaroff-Riley II sources (Snellen et al.
2003). In the frustration scenario, the radio source is confined and
frustrated by a cocoon of dense gas and dust for its entire life.
However, the far-infrared fluxes and near-infrared propertities of
CSS sources are roughly consistent with those of radio sources with
extended jets, which suggest that they have similar dust content in
the circumnuclear environment (e.g., Fanti et al. 2000; de Vries et
al. 1998).

If the evolutionary hypothesis of the youth scenario is correct,
study of the early-phase of the radio activity and feedback in
CSS/GPS sources may impact our understanding of galaxy evolution.
There is much evidence to suggest that, in a significant fraction of
the radio source population, the radio sources are triggered by
merger events of two or more galaxies (e.g., Heckman et al. 1986).
There are many compact radio sources exhibiting features attributed
to mergers, such as double nuclei, tidal tails, arcs of emission and
distorted isophotes, suggesting that these sources are possibly
observed relatively shortly after the merger event (Stanghellini et
al. 1993). The merger of galaxies will deposit large quantities of
gas and dust into the nuclear region, the central black hole (BH)
will grow rapidly through the merger-induced accretion, and
eventually form strong winds/outflows during fast accretion (e.g.,
Di Matteo et al. 2005). Narrow line Seyfert 1 galaxies (NLS1) are a
particular class of active galactic nuclei (AGN). The most widely
accepted paradigm for NLS1s is that they accrete at close to the
Eddington rate and have smaller BH masses compared broad line AGN,
which might be in the early stage of AGN evolution (e.g., Mineshige
et al. 2000; Boroson 2002; Grupe 2004; Bian \& Zhao 2004). Most of
the radio loud NLS1s galaxies are compact, steep-spectrum sources,
which are similar to CSS/GPS sources (e.g., Gallo et al. 2006;
Komossa et al. 2006). Therefore, compact radio sources (CSS/GPS) may
not only be in the early stage of radio activity, but also in the
early stage of BH accretion.

The mass of supermassive BH in the AGN is a key parameter to
understand the nuclear energy generation as well as BH formation and
evolution. There are several methods for estimating the BH masses in
AGN. In quiescent galaxies, dynamical modeling of either stellar
kinematics (e.g., Kormendy 1988) or gas motions (e.g., Harms 1994)
is used to determine central BH masses (e.g., Kormendy \& Gebhardt
2001 for a review). The correlation between the bulge stellar
velocity dispersion $\sigma_{*}$ and the central BH mass $M_{\rm
bh}$ is now well established in nearby inactive galaxies ($M_{\rm
bh}-\sigma_{*}$ relation, e.g., Gebhardt et al. 2000a; Ferrarese \&
Merritt 2000; Tremaine et al. 2002). Interestingly, the $M_{\rm
bh}-\sigma_{*}$ relation for normal galaxies also extends to AGN
(e.g., Gebhardt et al. 2000b). Unfortunately, the bulge stellar
velocity dispersion in bright AGN is generally difficult to measure
directly. In order to derive the BH mass in a larger sample of AGN,
the width of the [O III] line emitted by the narrow-line region is
frequently used as a surrogate for the stellar velocity dispersion,
since the kinematics of [O III] emission line is believed to be
dominated by the gravitational potential of the bulge. Nelson \&
Whittle (1996) found $\sigma_{\rm [O\ III]}\simeq\sigma_{*}$ in a
sample of 66 Seyfert galaxies, where $\sigma_{\rm [O\ III]}=\rm FWHM
(O\ III)/2.35$ and FWHM is full width at half-maximum of the
emission line. Nelson (2000) and Boroson (2003) found that $M_{\rm
bh}$ and $\sigma_{\rm [O\ III]}$ are strongly correlated and
consistent with the Tremaine et al. (2002) relation, although with
substantial scatter. The reverberation mapping (RM) method provides
a virial estimate of the BH mass in type 1 AGN. Using this
technique, the size of the broad line region (BLR) can be measured
using the time lag between the variations of the continuum and
emission line fluxes. The BH mass can then be estimated with $M_{\rm
bh}\sim R_{\rm BLR}V^{2}/G$, if we assume the gas near the BH is
virialized, where $R_{\rm BLR}$ is the BLR size, $V$ is the
characteristic velocity (e.g.,FWHM), and $G$ is the gravitational
constant (e.g., Blandford \& McKee 1982; Peterson 1993; Peterson
2004, and references therein). The most important result of the RM
method is the discovery of a simple scaling relationship between the
BLR size and the optical luminosity, which provides a useful
secondary method of mass determination (e.g., Wandel et al. 1999;
Kaspi et al. 2000; Vestergaard 2002; Kaspi et al. 2005; Greene \& Ho
2005a; Kong et al. 2006). This technique has been widely used in the
literature (e.g., Gu et al. 2001; Wu et al. 2004).

Despite the widespread use of $\sigma_{\rm [O\ III]}$ as a surrogate
of $\sigma_{*}$ in AGN, there are also several exceptions. In
particular, one should be cautious where the [O III] kinematics may
be affected by other drivers (e.g., outflows/winds) apart from the
first driver of gravitational potential of bulge. For example, it is
found that NLS1s as a class deviate systematically from the $M_{\rm
bh}-\sigma_{*}$ relation seen in normal galaxies, when assuming
$\sigma_{*}=\sigma_{\rm [O\ III]}$ (e.g., Grupe \& Mathur 2004; Bian
\& Zhao 2004). However, Barth et al. (2005) measured $\sigma_{*}$ in
19 NLS1s, and found that these NLS1s still follow the $M_{\rm
bh}-\sigma_{*}$ relation. The width of low-ionization lines (e.g.,
[S II]$\lambda6716,6731$) or the core component of [O III] (after
removal of asymmetric blue wings) of NLS1s is still a good proxy for
stellar velocity dispersion (e.g., Greene \& Ho 2005b; Komossa et
al. 2008). Therefore, the accretion activity (e.g., accretion-driven
outflows/winds) may affect the kinematics of the [O III] emission
line (e.g., Greene \& Ho 2005b; Bian et al. 2005; Komossa et al.
2008). The other possible exception is the [O III] narrow emission
line in young radio galaxies (CSS/GPS sources) and some luminous
linear radio sources, which, normally, have highly broadened line
profiles (e.g., Gelderman \& Whittle 1994; Nelson \& Whittle 1996;
Tadhunter et al. 2001; O'Dea et al. 2002; Holt et al. 2003, 2006,
2008). High resolution optical and radio imaging show that the [O
III] optical narrow emission line is strongly aligned with each
other in CSS sources at all redshifts (e.g., de Vries et al. 1997;
de Vries et al. 1999; Axon et al. 2000; Privon et al. 2008). The
expanding radio jets provide a convenient driving mechanism for the
observed nuclear outflows which may broaden the [O III] emission
line (jet-driven outflows, e.g., Tadhunter et al. 2001; Holt et al.
2003, 2006, 2008). Labiano (2008) also found that the [O III]
luminosity is well correlated with the CSS/GPS source size, which
seems to support the jet-driven scenario. The ionization diagnostics
on different narrow lines found that mix of shock and
photo-ionization give the best explainations in the extended
emission line gas (e.g., Inskip et al. 2002; Labiano et al. 2005;
Holt et al. 2006; Humphrey et al. 2008).

 Much of the work on the compact radio galaxies has
concentrated on the radio wavebands and optical emission lines, but
relatively little attention has been paid to their central engine
properties (e.g., BH masses and accretion modes). In particular, the
accretion-driven outflows/winds may also affect the kinematics of
the [O III] emission line in the radio loud CSS/GPS sources in
addition to jet-driven outflows, if their Eddington ratios are high
as those of radio quiet NLS1s. Therefore, it is important to test
these two scenarios in young radio galaxies. In this paper, we
present the BH masses, Eddington ratios, and the possible relations
between the [O III] kinematics and the accretion/jet properties for
a sample of CSS/GPS sources. Throughout this paper, we assume the
following cosmology: $H_{0}=70\ \rm km\ s^{-1} Mpc^{-1}$,
$\Omega_{0}=0.3$ and $\Omega_{\Lambda}=0.7$.

\begin{table*}
  \centerline{\bf Table 1. BH masses, Eddington ratios, [O III] kinematics, and jet
  powers.}
  \begin{tabular}{lcccclllcccc}\hline
Source & z$^a$ & Refs. & $M_{\rm bh}$ & $\frac{L_{\rm bol}}{L_{\rm
Edd}}$ & Method 1$^b$ & Method 2$^c$ & Refs.$^d$ & $\sigma_{[\rm O\
III]}$ & Refs. & $Q_{\rm
jet}$& Band\\
  &    &   & log ($\msun$) & log  &  &   &   & log (km/s) &   & log $\rm(erg/s)$ &
 \\
 (1) & (2) & (3) & (4) & (5) &(6) & (7) & (8) & (9)& (10) &
(11) & (12)
\\ \hline
                 &       &      &      &        &        CSS sample              &                 &         &      &       &      &  \\
                &       &      &      &        &                                 &                 &         &      &       &      &  \\
3C 43            & 1.459 & S89 & 9.2  & -1.0   & $L_{\rm opt}$+$V$                 & L5100             & M06     &  ... & ...   & 47.20 & 160M \\
3C 48            & 0.369 & S89 & 8.8  & -0.30  &$L_{\rm H\beta}$+$V(\rm H\beta)$   &$L_{\rm H\beta}$   & J91,C97 & 2.84 & G94   & 46.49 & 178M \\
3C 67            & 0.310 & S89 & 8.1  & -1.52  & $L_{\rm opt}$+$V$                 & L5100             &  S07    & 2.41 & G94   & 45.71 & 160M \\
3C 93.1          & 0.244 & S89 & 7.1  & ...    & $M_{\rm bh}$-$\sigma_{*}$         & Fitting           & W02     & 2.42 & G94   & 45.49 & 160M \\
3C 138           & 0.759 & S89 & 8.7  & -0.25  & $L_{\rm opt}$+$V$                 & Fitting           & W02     & 2.72 & G94   & 46.71 & 160M \\
3C 147           & 0.545 & S89 & 8.7  & -0.90  &$L_{\rm H\beta}$+$V(\rm H\beta)$   & $L_{\rm H\beta}$  & L96,M06 & 2.71 & G94   & 46.85 & 178M \\
3C 186           & 1.063 & S89 & 8.9  & -0.30  &$L_{\rm H\beta}$+$V(\rm H\beta)$   & $L_{\rm H\beta}$  & H03     & 2.67 & H03   & 46.88 & 151M\\
3C 190           & 1.195 & S89 & 7.8  & -0.13  &$L_{\rm H\beta}$+$V(\rm H\beta)$   & $L_{\rm H\beta}$  & H03     & 2.61 & H03   & 47.07 & 190M \\
3C 191           & 1.956 & S89 & 9.7  & -1.27  & $L_{\rm CIV}$+$V(\rm CIV)$        & $L_{\rm CIV}$     & M06,C91 & ...  &  ...  & 47.49 & 160M \\
3C 213.1         & 0.194 & S89 & 9.1  & -0.42  & $M_{\rm bh}$-$L_{\rm ion}$        & $L_{\rm ion}$$^e$     & D03     & 2.52 & G94 & 45.07 & 160M \\
3C 216           & 0.670 & S89 & 7.0  & -0.30  &$L_{\rm H\beta}$+$V(\rm H\beta)$   & $L_{\rm H\beta}$  & L96     & ...  &  ...  & 46.63 & 178M \\
3C 241           & 1.617 & S89 & 7.8  &  0.56  &$L_{\rm H\beta}$+$V(\rm H\beta)$   & $L_{\rm H\beta}$  & H03     & 2.63 & H03   & 47.71 & 151M\\
3C 268.3         & 0.372$^*$& S89 & 7.8  & ...    & $M_{R}$-$M_{\rm bh}$              & ...               & NED     & 2.46 & G94   & 45.87 & 178M\\
3C 277.1         & 0.320 & S89 & 7.6  & -0.14  &$L_{\rm H\beta}$+$V(\rm H\beta)$   & $L_{\rm H\beta}$  & G94     & 2.34 & G94   & 45.65 & 178M\\
3C 286           & 0.849 & S89 & 8.5  & -0.09  & $L_{\rm opt}$+$V$                 & L5100             & M06     & 2.31 & G94   & 46.87 & 151M\\
3C 287           & 1.055 & S89 & 9.6  & -0.85  & $L_{\rm opt}$+$V$                 & L5100             & M06     & ...  &  ...  & 46.87 & 151M \\
3C 293           & 0.045 & S04 & 8.0  & -1.78  & $M_{\rm bh}$-$\sigma_{*}$         & L5100             & W02,X08 & ...  &  ...  & 44.22 & 151M \\
3C 298           & 1.439 & S89 & 10.4 & -0.83  & $L_{\rm opt}$+$V$                 & L5100             & M04     &  ... &  ...  & 47.68 & 178M\\
3C 303.1         & 0.270$^*$ & S89 & 8.4  &  ...   & $M_{R}$-$M_{\rm bh}$              & ...               & NED     & 2.54 & G94   & 45.54 & 151M \\
3C 309.1         & 0.905 & S89 & 9.1  & -0.45  &$L_{\rm H\beta}$+$V(\rm H\beta)$   & $L_{\rm H\beta}$  & L96     &  ... &  ...  & 46.95 & 178M\\
3C 343           & 0.988 & S89 & 7.5  &  0.14  & $L_{\rm opt}$+$V$                 & Fitting           & W02     &  ... &  ...  & 46.81 & 178M\\
3C 346           & 0.162 & S89 & 8.8  & -2.54  & $M_{R}$-$M_{\rm bh}$              & L5100             & NED,M04 & 2.60 & G94   & 45.16 & 160M\\
3C 380           & 0.692 & S89 & 9.4  & -1.15  & $L_{\rm opt}$+$V$                 & L5100             & M04     & 2.41 & G94   & 47.07 & 178M\\
3C 454           & 1.757 & S89 & 8.6  & -0.58  & $L_{\rm CIV}$+$V(\rm CIV)$        & $L_{\rm CIV}$     & B90     & ...  &  ...  & 47.25 & 160M \\
3C 455           & 0.543 & S89 & 8.5  & -1.62  & $L_{\rm opt}$+$V$                 & L5100             & M04     & 2.25 & G94   & 45.84 & 160M\\
3C 459           & 0.220 & H08 & 8.5  &   ...  & $M_{R}$-$M_{\rm bh}$              & ...               & NED     & 2.63 & H08   & 45.79 & 160M \\
2342+821         & 0.735 & S89 & 7.5  & 0.18   & $L_{\rm opt}$+$V$                 & Fitting           & W02     & 2.39 & G94   & 46.00 & 151M \\
4C 14.82         & 0.235 & S89 & 7.9  & -0.79  &$L_{\rm H\beta}$+$V(\rm H\beta)$   & $L_{H\beta}$      & C97     & 2.33 & G94   & 45.20 & 160M \\
PKS 0252-71      & 0.563$^*$ & H08 & 8.0  &   ...  & $M_{R}$-$M_{\rm bh}$              & ...               & NED     & 2.48 & H08   & 46.11 & 408M \\
PKS 1151-348     & 0.258 & M97 & 9.1  & -1.54  & $L_{\rm opt}$+$V$                 & Fitting           & W02     & ...  &  ...  & 45.53 & 160M \\
PKS 1221-423     & 0.171 & J05 & 7.8  &  ...   & $M_{R}$-$M_{\rm bh}$              & ...               & NED     & 2.53 & J05   & 45.04 & 160M \\
PKS 1524-136     & 1.687 & E04 & 9.1  & -0.91  & $M_{R}$-$M_{\rm bh}$              & L5100             & P06     &...   & ...   & 47.11 & 160M \\
PKS 1549-79      & 0.152 & H08 & 8.0  & -0.32  & $M_{R}$-$M_{\rm bh}$              & $L_{H\beta}$      & H06     & 2.77 & H06   & 45.29 & 408M \\
PKS 2004-447     & 0.240 & G06 & 7.6  & -0.26  & $L_{\rm opt}$+$V$                 & Fitting           & W02     & ...  &  ...  & 44.94 & 408M \\
SDSS 1722+565    & 0.425 & K06 & 7.5  & -0.14  & $L_{\rm opt}$+$V$                 & L5100             & K06     & 2.32 & K06   & 44.48 & 330M \\
RX J0134-4258    & 0.237 & K06 & 7.6  & 0.01   & $L_{\rm opt}$+$V$                 & L5100             & K06     & 2.34 & K06   & 44.73 &4850M \\
TeX 11111+329    & 0.189 & K06 & 7.3  & -0.21  & $L_{\rm opt}$+$V$                 & L5100             & K06     & 2.80 & K06   &  ...  & ...  \\
IRAS 2018-2244   & 0.185 & K06 & 6.5  & 1.25   & $L_{\rm opt}$+$V$                 & L5100             & K06     & 2.47 & K06   & 44.0 & 352M\\
                 &       &      &      &        &                                &                &         &      &       &       &  \\
                 &       &      &      &        & GPS sample                     &                &         &      &       &       &  \\
                 &       &      &      &        &                                &                &         &      &       &       &  \\
B 0019-0001      & 0.305 & S03 & 8.4  &   ...  & $M_{\rm bh}$-$\sigma_{*}$      &...             & S03    & ...   & ...   & 45.09 & 178M \\
B 0830+5813      & 0.094 & S03 & 7.6  &   ...  & $M_{\rm bh}$-$\sigma_{*}$      &...             & S03    & ...   & ...   & 42.84 & 325M \\
B 0941-0805      & 0.228 & S03 & 7.7  &   ...  & $M_{\rm bh}$-$\sigma_{*}$      &...             & S03    & ...   & ...    &   ... & ... \\
B 1819+6707      & 0.221 & S03 & 8.4  &   ...  & $M_{\rm bh}$-$\sigma_{*}$      &...             & S03    &$<$2.2 & S99 & 44.25 & 325M \\
B 1144+3517      & 0.063 & S03 & 8.2  &   ...  & $M_{\rm bh}$-$\sigma_{*}$      &...             & S03    & ...   & ...   & 42.93 & 151M \\
B 1946+7048      & 0.101 & S03 & 7.9  &   ...  & $M_{\rm bh}$-$\sigma_{*}$      &...             & S03    &$<$2.15& S99 & 43.56 & 325M \\
B 2352+4933      & 0.238 & S03 & 8.2  &  ...   & $M_{\rm bh}$-$\sigma_{*}$      &...             & S03    &  ...  & ...   & 44.65 & 151M \\
B 0537+6444      & 2.417 & S99 & 7.7  &  -0.29 & $L_{\rm CIV}$+$V(\rm CIV)$     & $L_{\rm CIV}$  & S99    &  ...  & ...   & 45.39 & 325M \\
B 0544+5847      & 2.860 & S99 & 8.0  &  -0.91 & $L_{\rm CIV}$+$V(\rm CIV)$     & $L_{\rm CIV}$  & S99    &  ...  & ...   & 45.77 & 325M \\
B 0601+5753      & 1.840 & S99 & 7.9  & -0.53  & $L_{\rm CIV}$+$V(\rm CIV)$     & $L_{\rm CIV}$  & S99    & ...   &  ...  & 45.16 & 325M \\
B 0755+6354      & 3.005 & S99 & 8.3  &  -0.41 & $L_{\rm CIV}$+$V(\rm CIV)$     & $L_{\rm CIV}$  & S99    & ...   &  ...  & 45.72 & 325M \\
B 0758+5929      & 1.977 & S99 & 8.1  & -1.05  & $L_{\rm CIV}$+$V(\rm CIV)$     & $L_{\rm CIV}$  & S99    & ...   &  ...  & 45.84 & 325M \\
B 0826+7045      & 2.008 & S99 & 8.5  &  -0.93 & $L_{\rm CIV}$+$V(\rm CIV)$     & $L_{\rm CIV}$  & S99    &  ...  & ...   & 45.46 &325M \\
B 1538+5920      & 3.880 & S99 & 7.7  &  -0.16 & $L_{\rm CIV}$+$V(\rm CIV)$     & $L_{\rm CIV}$  & S99    &  ...  & ...   & 45.99 & 325M \\
B 1550+5815      & 1.320 & S99 & 8.9  &  -0.65 &$L_{\rm MgII}$+$V(\rm MgII)$    &$L_{\rm MgII}$  & S99    &  ...  & ...   & 45.42 & 325M \\
B 1622+6630      & 0.202 & S99 & 7.7  &  -0.94 &$L_{\rm H\alpha}$+$V(\rm H\alpha)$& $L_{\rm H\alpha}$ & S99   &$<$2.15  & S99 & 43.26 & 325M \\
B 1642+6701      & 1.895 & S99 & 9.1  & -0.79  & $L_{\rm CIV}$+$V(\rm CIV)$     & $L_{\rm CIV}$  & S99    &  ... & ...   & 45.89 & 325M \\
B 1647+6225      & 2.180 & S99 & 7.2  & -0.38  & $L_{\rm CIV}$+$V(\rm CIV)$     & $L_{\rm CIV}$  & S99    &  ... & ...   & 45.50 & 325M \\
B 1746+6921      & 1.886 & S99 & 7.5  &  -0.22 & $L_{\rm CIV}$+$V(\rm CIV)$     & $L_{\rm CIV}$  & S99    &  ... & ...   & 45.65 & 325M \\

\hline
\end{tabular}
\end{table*}
\begin{table*}
\end{table*}

\begin{table*}
  \centerline{\bf Table 1. (Continued)}
  \begin{tabular}{lcccclllcccc}\hline
Source & z & Refs. & $M_{\rm bh}$ & $\frac{L_{\rm bol}}{L_{\rm
Edd}}$ & Method 1 & Method 2 & Refs. & $\sigma_{[\rm O\ III]}$ &
Refs. & $Q_{\rm
jet}$& Band\\
  &    &   & log ($\msun$) & log  &  &   &   & log (km/s) &   & log $\rm(erg/s)$ &
 \\
 (1) & (2) & (3) & (4) & (5) &(6) & (7) & (8) & (9)& (10) &
(11) & (12)
\\ \hline
B 1945+6024      & 2.700 & S99 & 7.5  &  -0.70 & $L_{\rm CIV}$+$V(\rm CIV)$     & $L_{\rm CIV}$  & S99    &  ... &  ...  & 45.61 & 325M \\
B 1958+6158      & 1.820 & S99 & 7.8  &  -0.21 & $L_{\rm CIV}$+$V(\rm CIV)$     & $L_{\rm CIV}$  & S99    &  ... &  ...  & 45.53 & 325M \\
B 0237-233       & 2.223 & S99 & 9.5  & -0.77  & $L_{\rm CIV}$+$V(\rm CIV)$     & $L_{\rm CIV}$  & S99    &  ... &  ...  & 47.26 & 365M \\
B 0457+24        & 2.384 & S99 & 8.8  &  -0.77 & $L_{\rm CIV}$+$V(\rm CIV)$     & $L_{\rm CIV}$  & S99    &  ... &  ...  & 46.61 & 365M \\
B 2126-158       & 3.270 & S99 & 9.2  & -0.56  & $L_{\rm CIV}$+$V(\rm CIV)$     & $L_{\rm CIV}$  & S99    &  ... &  ...  & 46.30 & 333M \\
B 2134+004       & 1.936 & S99 & 9.1  &  -0.77 & $L_{\rm CIV}$+$V(\rm CIV)$     & $L_{\rm CIV}$  & S99    &  ... &  ...  & 46.88 & 365M \\
OQ 172           & 3.544 & O98 & 9.6  &  -0.10 & $L_{\rm H\beta}$+$V(\rm H\beta)$& $L_{\rm H\beta}$ & H03    & 2.97 & H03  & 47.61 & 408M \\
4C 12.50         & 0.124$^*$ & O98 & 7.8  &  ...   & $M_{\rm bh}$-$\sigma_{*}$      & ...            & D06    & 2.77 & G94  & 44.62 & 178M\\
\hline
\end{tabular}

\begin{minipage}{170mm}
 $References$: S89: Spencer et al. (1989);  H08: Holt, J. et al.
(2008); M97: Morganti et al. (1997);
 J05: Johnston et al. (2005); E04: Edwards \& Tingay (2004); G06: Gallo et al. (2006);
 K06: Komossa et al. (2006);  S03: Snellen et al. (2003); S99: Snellen et al. (1999);
 O98: O'Dea (1998); M06: McLure et al. (2006); J91: Jackson \& Browne (1991);
 C97: Corbin (1997); S07: Sikora et al. (2007); W02: Woo \& Urry (2002);
 L96: Lawrence et al. (1996); H03: Hirst et al. (2003); C91: Corbin (1991);
 D03: D'Elia et al. (2003); G94: Gelderman \& Whittle (1994); X08: Xie et al. (2008);
 M04: Marchesini et al. (2004); B90: Barthel, Tytler, \& Thomson (1990); P06: Peng et al. (2006);
 H06: Holt et al. (2006); D06: Dasyra et al. (2006).

 $Note$:\\
 (a) The redshifts are selected from NED$^1$, and the redshifts with stars are
 updated in Holt et al. 2008.\\
 (b) Methods for calculating the BH masses in column 4 as
 described in section 3.1.\\
 (c) Data used to calculate the bolometric luminosities in column 5 as
 described in section 3.2.\\
 (d) References for related data used in columns 6 and 7.\\
 (e) The BH mass is calculated from the relation between ionizing luminosity and BH
 mass (refer D'Elia et al. 2003 for more details).

 \end{minipage}
\end{table*}

\section{Sample}

  Our goal is to investigate the nuclear properties (e.g., accretion
and jet) and their possible relation to the kinematics of the [O
III] emission line in the compact radio galaxies (CSS/GPS sources),
whose selection is unbiased with respect to the nuclear properties
themselves. For each source, we must be able to estimate the BH
mass, and/or bolometric luminosity, jet power, the [O III]
kinematics (e.g., FWHM). We present 38 CSS sources, of which 27
sources are selected from Spencer et al. (1989) and the other 11
sources are selected from different papers (see Table 1 for
references). We also selected 27 GPS sources from the literature
from various surveys (e.g., Snellen et al. 1999, 2003), of which two
sources (OQ 172 and 4C 12.50) were defined as CSS sources in Spencer
et al. (1989), but later were defined to be GPS sources according to
their radio spectra and sizes (e.g., O'Dea 1998 and references
therein). Therefore, our sample includes 65 compact radio sources
(38 CSS + 27 GPS objects). We note that our sample is a
characteristic, rather than complete sample of compact radio
sources. The histogram distributions of the redshift for our sample
are shown in Figure 1. The dotted-line is for CSS sources, the
dashed-line is for GPS sources, and the solid-line is their sum.
\begin{figure}
\centerline{\psfig{figure=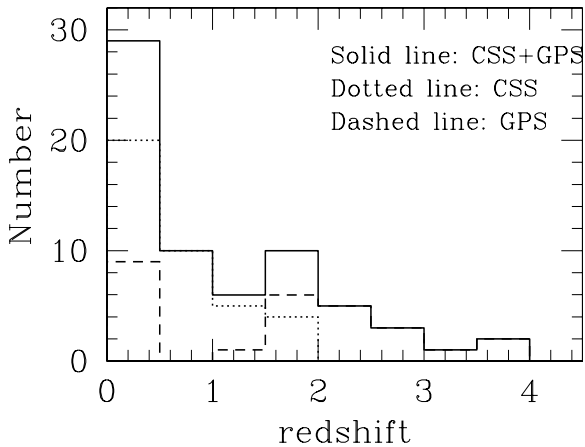,width=8.5cm,height=6.5cm}}
\caption{The histogram distribution of redshift. }
\end{figure}

 Table 1 lists our sample with the relevant information. Columns
 (1)-(3) give the object's name, redshift, and references for the
classification as CSS/GPS sources. Columns (4)-(8) list the BH
 masses, Eddington ratios, methods for deriving the BH masses,
 methods for deriving bolometric luminosities, and references for
 related information in columns 6 and/or 7. Columns (9)-(10) give $\sigma_{\rm [O\ III]}$ and
 their references. Columns (11)-(12) list the jet power obtained from NED\footnote{NASA/IPAC Extragalactic Database;
  http://nedwww.ipac.caltech.edu}, and the waveband in which the power was calculated.

\begin{figure*}
\begin{center}
\includegraphics[width=16cm]{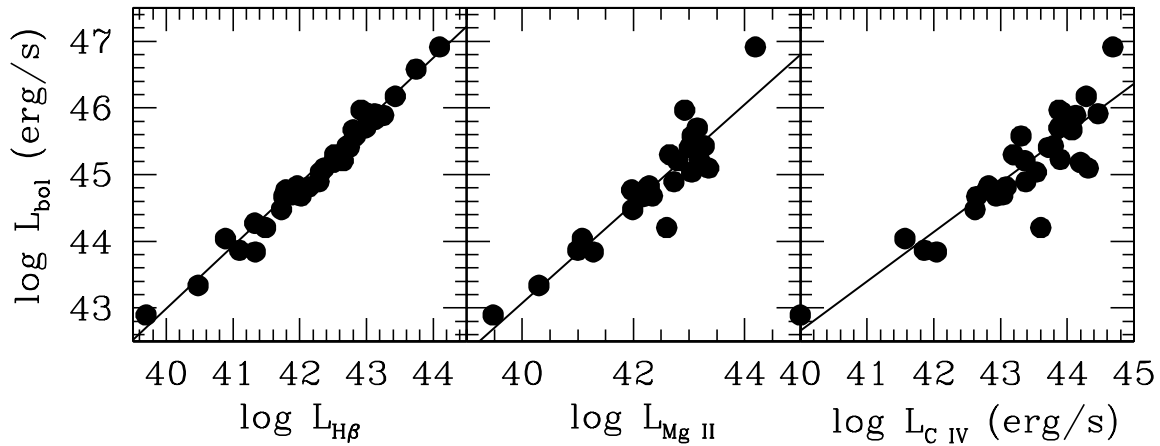}
\caption{The correlations between the bolometric luminosities and
broad line luminosities, $H\beta$ (left), Mg II (middle), and C IV
(right) respectively for a sample of radio quiet AGN from Kaspi et
al. 2005, where the bolometric luminosity $L_{\rm bol}=9\lambda
L_{\lambda} (5100\rm \AA)$ is assumed (e.g., Kaspi et al. 2000). The
solid lines are their best fittings. The optical continuum 5100$\rm
\AA$ luminosities and $\rm H\beta$ luminosities are selected from
Kaspi et al. 2005, while Mg II, C IV luminosities are selected from
Kong et al. 2006. }
\end{center}
\end{figure*}

\section{Methods}

\subsection{BH mass estimation}
For the RM method, it takes a long time to simultaneously monitor
the variability of the broad emission line and the continuum, and
then obtain the BLR size. Fortunately, Kaspi et al. (2000) found an
empirical correlation between the BLR size and the monochromatic
luminosity at 5100$\rm \AA$. Recently, Kaspi et al. (2005)
reinvestigated the relationship between the characteristic BLR size,
$R_{\rm BLR}$, and the $\rm H\beta$ emission line, X-ray, UV and
optical continuum luminosities. Several other empirical relations
between $R_{\rm BLR}$ and emission line luminosities (e.g., $\rm
H\alpha$, C IV and Mg II) have also been investigated (e.g.,
Vestergaard 2002; Greene \& Ho 2005a; Kong et al. 2006; McGill et
al. 2008). The BH mass then can be estimated with the assumption of
virialized gas around the BH, \be M_{\rm bh}=R_{\rm BLR}V^{2}G^{-1},
\ee where the quantity $V$ can be estimated from the FWHM of the
broad emission lines ($V=fV_{\rm FWHM}$), and $f=\sqrt{3}/2$
assuming an isotropic distribution of the BLR clouds (e.g., Wandel
et al. 1999; Kaspi et al. 2000).

The optical continuum in the radio loud (RL) AGN may be contaminated
by the nonthermal emission from a jet: their optical luminosity and
hence the viral BH mass may be over-estimated. In general,
steep-spectrum radio sources tend to be oriented with jets pointing
away from our line of sight, and the jet contribution at optical
waveband may be negligible due to the weakness of beaming effects.
Fanti et al. (1990) proposed that CSS sources were not strongly
affected by Doppler boosting. Woo \& Urry (2002) found that the BH
masses derived from the optical luminosities and FWHM of the broad
emission lines in RL AGN are consistent with those derived from
other independent methods. Therefore, BH masses derived from broad
line width and optical continuum/line luminosities for CSS sources
should be credible. However, Baker \& Hunstead (1995) found that the
composite GPS quasar spectrum is clearly different from that of CSS
quasars, but similar to that of flat-spectrum radio quasars (FSRQs).
If this is true, the optical emission in GPS sources may be strongly
contaminated by the synchrotron emission from the jets, as in FSRQs
(e.g., Liu et al. 2006). Therefore, we use broad line luminosities
to calculated the BH masses for GPS sources in our sample. We list
the equations used to estimate BH mass from the broad emission line
($\rm H\alpha$ and $\rm H\beta$ from Greene \& Ho 2005a; Mg II and C
IV from Kong et al. 2006),

\be M_{\rm bh}=2.4\times10^{6} \left(\frac{L_{\rm H\beta}}{10^{42}\
\rm ergs\ s^{-1}} \right)^{0.59} \left(\frac{\rm FWHM_{\rm
H\beta}}{10^{3}\ \rm km\ s^{-1}} \right)^{2} \msun, \ee

\be M_{\rm bh}=1.3\times10^{6} \left(\frac{L_{\rm H\alpha}}{10^{42}\
\rm ergs\ s^{-1}} \right)^{0.57} \left(\frac{\rm FWHM_{\rm
H\alpha}}{10^{3}\ \rm km\ s^{-1}} \right)^{2.06} \msun, \ee

\be M_{\rm bh}=2.9\times10^{6} \left(\frac{L_{\rm Mg II}}{10^{42}\
\rm ergs\ s^{-1}} \right)^{0.57} \left(\frac{\rm FWHM_{\rm Mg
II}}{10^{3}\ \rm km\ s^{-1}} \right)^{2} \msun, \ee

\be M_{\rm bh}=4.6\times10^{5} \left(\frac{L_{\rm C IV}}{10^{42}\
\rm ergs\ s^{-1}} \right)^{0.60} \left(\frac{\rm FWHM_{\rm C
IV}}{10^{3}\ \rm km\ s^{-1}} \right)^{2} \msun. \ee BH masses for
some CSS sources, estimated from the optical luminosity and broad
line width, are also selected from the literature directly if there
is no line luminosity, but we correct the optical luminosity to our
cosmology (see Table 1).

The BH mass can also be derived from the $M_{\rm bh}-\sigma_{*}$
relation (e.g., Tremaine et al. 2002) for nearby AGN (e.g., Gebhardt
et al. 2000b), \be M_{\rm bh}=10^{8.13}\left(\frac{\sigma_{*}}{200\
\rm km\ s^{-1}} \right)^{4.02} \msun, \ee with an intrinsic scatter
$\simeq0.3$ dex .

Alternatively, the BH mass can be estimated from the galactic host
bulge luminosity. McLure \& Dunlop (2002) found that the BH mass is
tightly correlated with the luminosity of the host galaxy for a
sample of both active and inactive galaxies, with a scatter $\simeq
0.4$ dex. We use the relation between the host galaxy absolute
magnitude $M_{R}$ at $R$-band and BH mass proposed by McLure et al.
(2004), \be \log_{10}(M_{\rm bh}/\msun)=-0.5M_{R}-2.74,\ee to
estimate the central BH masses for the sources without data for
broad emission lines or bulge stellar velocity dispersion.

\subsection{Bolometric luminosity estimation}

  The Bolometric luminosity of AGN is sometimes approximated from the
  optical luminosity since integration of the spectral energy
  distribution (SED) is usually hampered by the lack of
  multi-wavelength observational data that spans many decades in
  wavelength. It is found that the bolometric luminosity is roughly 10 times
  the optical luminosity, $L_{\rm bol}\simeq 10 L_{\rm opt}$,
  where bolometric luminosity was integrated for the SEDs of AGN (e.g.,
   Kaspi et al. 2000; Woo \& Urry 2002; see also Vasudevan \& Fabian
  2007). In this work, we adopt $L_{\rm bol}=9 L_{5100}$,
  where $L_{5100}=\lambda L_{\lambda}(5100\rm \AA)$ (Kaspi et al.
  2000). However, the monochromatic optical luminosity cannot be
  used to derive the bolometric luminosity if beaming is
  important in radio loud AGN (e.g., BL Lac, FSRQs and maybe also GPS sources in our
  work). We know that the broad line emission is illuminated by the ionizing luminosity from the central engine,
  and, therefore, it can be used as a good indicator of thermal
  optical emission. We can estimate the thermal optical
  emission using the broad emission lines for the sources in which the optical emission may be
  contaminated by the jet (BL Lacs, FSRQs, and/or GPS sources). To do so, we
  fit the bolometric luminosities (assuming $L_{\rm bol}=9 L_{5100}$)
  and the broad line luminosities ($\rm H\beta$, Mg II and C IV) for a sample of radio quiet AGN in Kaspi et al.
  (2005), where the optical emission is believed to be free
  from contamination by the nonthermal synchrotron emission (the jet
  is weak, if present).  The $\rm H\beta$ luminosities and 5100$\rm \AA$
   monochromatic luminosites are selected from Kaspi et al.
  (2005) for 35 AGN, of which 27 sources with Mg II luminosities,
  and 32 sources with C IV luminosities are selected from Kong et
  al. (2006). The relation between the bolometric luminosities
  and different broad emission lines ($\rm H\beta$, Mg II and C IV)
  are shown in Figure 2, and their linear regressions are
  \be \log L_{\rm bol}=0.94\pm0.03 \log L_{\rm H\beta}+5.39\pm1.14, \ee

\be \log L_{\rm bol}=0.74\pm0.07 \log L_{\rm MgII }+13.06\pm3.02,\ee

\be \log L_{\rm bol}=0.74\pm0.06 \log L_{\rm CIV}+13.31\pm2.61. \ee
With Eqs. 8-10, we can calculate the bolometric luminosities for the
radio loud AGN from their broad emission line luminosities,
especially for sources that are probably significantly contaminated
by the beamed emission from the relativistic jets. We note that our
bolometric luminosities derived from the $\rm H\beta$ luminosities
are roughly 3 times larger than that derived using the popular
relation between the bolometric luminosity and the BLR luminosity
$L_{\rm bol}\simeq10 L_{\rm BLR}$, where $L_{\rm
BLR}\simeq25.26L_{\rm H\beta}$ (e.g., Celotti et al. 1997; Liu et
al. 2006 and references therein).

\subsection{Jet power estimation}
Jet power is a fundamental parameter reflecting the energy transport
to large spatial scales from the central engine by the radio jet.
The jet power can be estimated from the low frequency radio
emission, which is usually emitted from the extended optically thin
radio lobes, and is free from Doppler beaming due to the low bulk
velocity of the lobe motion (e.g., Willott et al. 1999). In this
work, we use the formula proposed by Willott et al. (1999),
 \be Q_{\rm
jet}\simeq3\times10^{38}f^{3/2}L151^{6/7} (\rm W),\ee where
$L_{151}$ is the radio luminosity from the lobes measured at 151 MHz
in units of $10^{28}\rm W\ Hz^{-1}sr^{-1}$. Willott et al. (1999)
have argued that the normalization is uncertain and introduced the
factor $f$ ($1\leq f \leq 20$) to account for these uncertainties.
Blundell \& Rawlings (2000) argued that $f\simeq10$ is a likely
consequence of the evolution of magnetic field strengths as radio
sources evolve. In this work, we adopt $f=10$, and search the low
frequency radio data from NED for CSS/GPS sources. The extended
emission was $K-$corrected to 151 MHz assuming $\alpha=1$
($F_{\nu}\propto\nu^{-\alpha}$), and the jet power was then
calculated from Eq. 11 (see Table 1).

\section{Results}

\subsection{BH mass, Eddington ratio and jet power}

The BH masses of 65 compact radio galaxies (38 CSS+27 GPS objects)
are calculated in this work, or collected from the literature, using
several different methods as described in Sect. 3.1. We find that
the BH masses vary by several orders from
$10^{6.5}$-$10^{10.5}\msun$, for both CSS and GPS sources. The
histogram distributions of the BH masses are shown in Figure 3, the
dotted-line, dashed-line and solid-line are for CSS sources, GPS
sources, and the whole sample respectively. We find that the average
logarithmic BH masses are $<$$\log M_{\rm bh}$$>$=8.31, 8.23, 8.28
for 38 CSS sources, 27 GPS sources, and the whole sample
respectively.

\begin{figure}
\centerline{\psfig{figure=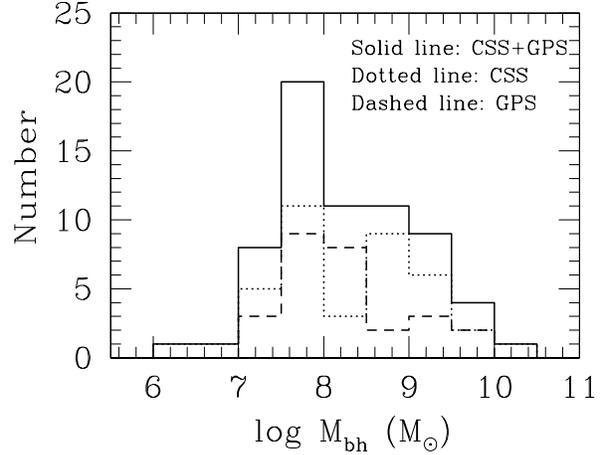,width=8.5cm,height=6.5cm}}
\caption{The histogram distribution of the black hole masses. }
\label{}
\end{figure}

\begin{figure}
\centerline{\psfig{figure=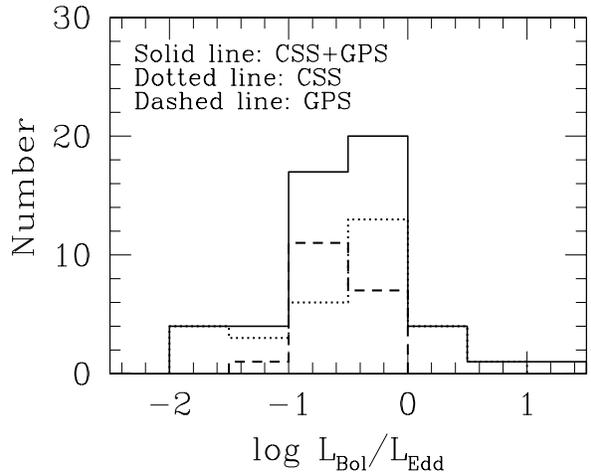,width=8.5cm,height=6.5cm}}
\caption{The histogram distribution of the Eddington ratios. }
\label{}
\end{figure}

The correlation between the bolometric luminosities and broad
emission line luminosities ($\rm H\beta$, Mg II and C IV) for a
sample of radio quiet AGN are explored in this work (see Figure 2
and Sect. 3.2 for more details). We then calculate the bolometric
luminosities for all GPS sources, and some of CSS sources from their
broad emission lines with the above empirical correlations (Eqs.
8-10), except for those with bolometric luminosities derived from
their SED fittings. The Eddington ratios are calculated with $L_{\rm
bol}/L_{\rm Edd}$, where $L_{\rm Edd}=1.38\times10^{38}M_{\rm
bh}/\msun$. We find that young radio galaxies (GPS/CSS sources) have
relatively high Eddington ratios, $10^{-2}$$<$ $L_{\rm bol}/L_{\rm
Edd}$$<$ several (see Figure 4). The average logarithmic Eddington
ratios are $<$$\log L_{\rm bol}/L_{\rm Edd}$$>$=-0.55, -0.58 and
-0.56 for the 32 CSS sources, the 19 GPS sources, and their sum,
respectively.

 It is found that the jet power ranges from
$\sim$$10^{42}$ to $\sim$$10^{47}\rm erg\ s^{-1}$ for both the CSS
and GPS sources. The histogram distributions of the jet power for
the CSS/GPS sources are shown in Figure 5. The average logarithmic
jet powers are $<$$\log Q_{\rm jet}$$>$=46.1 and 45.3 for the 37 CSS
and the 26 GPS sources respectively. The average jet power in the
CSS sources is larger than that of the GPS sources based on our
sample.

\begin{figure}
\centerline{\psfig{figure=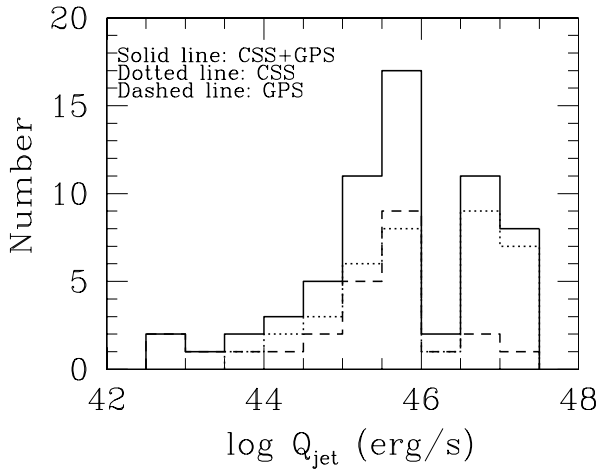,width=8.5cm,height=6.5cm}}
\caption{The histogram distribution of the jet power.} \label{}
\end{figure}

\subsection{[O III] kinematics and accretion/jet properties }

\begin{figure}
\centerline{\psfig{figure=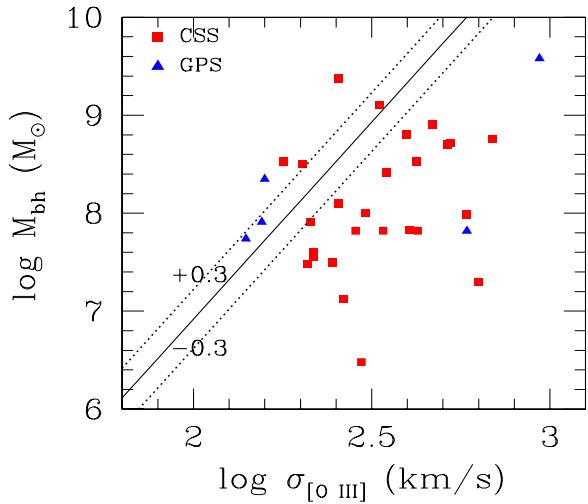,width=8.5cm,height=7.cm}}
\caption{The velocity dispersion $\sigma_{\rm [O\ III]}$ vs. BH mass
$M_{\rm bh}$. The solid line is the $M_{\rm bh}-\sigma_{*}$ relation
of Tremaine et al. 2002 for 31 nearby inactive galaxies, and the
dotted lines are their intrinsic scatter $\pm0.3$ dex. }
\label{mqj10}
\end{figure}

Figure 6 plots BH mass, $M_{\rm bh}$, versus velocity dispersion,
$\sigma_{\rm [O\ III]}$, derived from FWHM(O III), which includes 23
CSS (squares) and 5 GPS sources (triangles). The solid line is the
relation of Tremaine et al. (2002) obtained for 31 nearby inactive
galaxies. It is clear that the CSS sources systematically deviate
from the Tremaine et al. relation. Two of the five GPS sources also
deviate substantially (OQ 172 and 4C 12.50).

We calculate the deviation $\Delta\sigma\equiv\log \sigma_{\rm [O\
III]}-\log \sigma_{\rm [pred]}$ for our sample with the measured
$\sigma_{\rm [O\ III]}$, where $\sigma_{\rm [pred]}$ is calculated
from the Tremaine et al. relation (Eq. 6) with the estimated BH
mass. Figure 7a and 7b plot $\Delta\sigma$ versus jet power, $Q_{\rm
jet}$, and bolometric luminosity, $L_{\rm bol}$, respectively. Using
the Spearman rank correlation analysis, there are no significant
correlations between $\Delta\sigma$ and $Q_{\rm jet}$ ($r=0.12$ and
$P_{\rm null}$=0.51) or between $\Delta\sigma$ and $L_{\rm bol}$
($r=0.11$ and $P_{\rm null}$=0.59).

\begin{figure}
\centerline{\psfig{figure=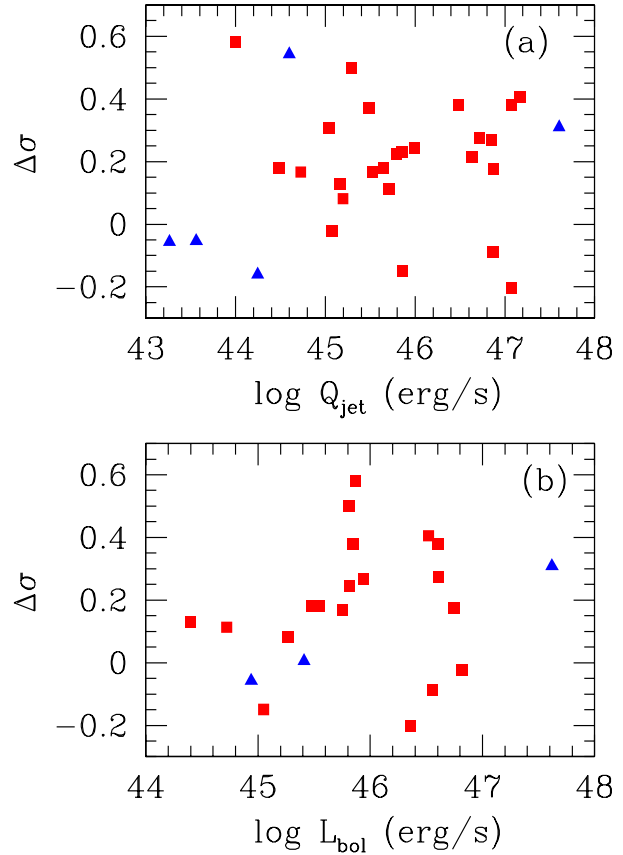,width=9.0cm,height=12.0cm}}
\caption{Jet power $Q_{\rm jet}$ vs. $\rm \Delta \sigma$ (top), and
accretion power $L_{\rm bol}$ vs. $\rm \Delta \sigma$ (bottom),
where $\rm \Delta \sigma \equiv log \sigma_{[O\ III]}-log
\sigma_{\rm [pred]}$.}
\end{figure}

Figure 8 plots $\Delta\sigma$ versus Eddington ratio $L_{\rm
bol}/L_{\rm Edd}$ for the compact radio galaxies (20 CSS + 3 GPS
sources). We find a significant correlation between $\Delta\sigma$
and $L_{\rm bol}/L_{\rm Edd}$ for these 23 young radio sources, with
the Spearman correlation coefficient $r=0.53$ and probability
$P_{\rm null}\simeq1\times10^{-3}$. For comparison with the radio
quiet AGN (the jet is weak, if present), we also calculate
$\Delta\sigma$ and the Eddington ratio $L_{\rm bol}/L_{\rm Edd}$ for
31 NLS1s (circles) and 49 QSOs (crosses). The NLS1s are selected
from Grupe \& Mathur (2004) and QSOs are selected from Shields et
al. (2003), where BH masses are calculated from $L_{5100}$ and the
broad line width, and bolometric luminosities are calculated from
$L_{5100}$. We find that the relation, $\Delta\sigma$-$L_{\rm
bol}/L_{\rm Edd}$, in the young radio galaxies (GPS/CSS) is similar
to that of radio quiet AGN, and the sources with $L_{\rm bol}/L_{\rm
Edd}$$\sim$1 have the largest deviations (see Figure 8).

\section{Discussion}

In this paper we investigate the central engine properties for a
sample of young radio galaxies (CSS/GPS sources), i.e., BH masses,
Eddington ratios, jet powers, and then explore the possible relation
between kinematics of [O III] narrow line and accretion/jet
activities.

The BH masses of the young radio galaxies are estimated from several
empirical relations in this work, as described in Sect. 3.1, and
some BH masses are selected from the literature after correcting to
our cosmology, if the luminosities are used in deriving BH masses.
We find that most of the BH masses are in the range of
$10^{7}$-$10^{10}\msun$ for both CSS and GPS sources (Figure 3),
which are typical BH masses for normal AGN. The mean BH mass of the
38 CSS sources is similar to that of the 27 GPS sources in our
sample, with no evident difference between them. More complete
sample is desired to further test this issue. However, the mean BH
mass of young radio galaxies (CSS/GPS sources, $<\log M_{\rm
bh}>\simeq$8.3) is systematically less than that of radio loud QSOs
($<\log M_{\rm bh}>\simeq$9.1, e.g., Marchesini et al. 2004) or low
redshift radio galaxies ($<\log M_{\rm bh}>\simeq$8.9, e.g., Bettoni
et al. 2003). One of the possible physical reasons is that the BHs
in young radio galaxies are still growing rapidly.

\begin{figure}
\centerline{\psfig{figure=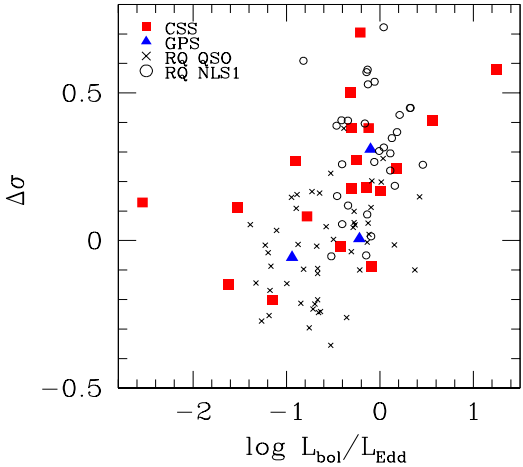,width=7.5cm,height=6.5cm}}
\caption{The relation between the Eddington ratio $L_{\rm
bol}/L_{Edd}$ and $\rm \Delta \sigma$. }
\end{figure}

The bolometric luminosities for young radio galaxies are estimated
from SED fitting, monochromatic optical luminosities or broad line
luminosities (see Sect. 3.2 for details). The bolometric
luminosities of some CSS/GPS sources are not estimated due to a lack
of information on their classification (e.g., type I or II) where
optical luminosities and then bolometric luminosities may be
underestimated if they are obscured by the putative torus. Normally,
steep-spectrum radio sources do not suffer from strong contamination
from the Doppler-boosted emission from the jet, due to the possible
large inclination angles. However, there is some evidence from radio
variability that at least some of the GPS sources are
Doppler-boosted (e.g., O'Dea 1998, and references therein). The
optical composite spectrum of some GPS sources is also clearly
different from that of CSS sources, which is similar to that of
FSRQs and may be contaminated by the nonthermal emission from jets
(e.g., Baker \& Hunstead 1995). Therefore, the bolometric
luminosities of all GPS sources are calculated from the broad line
luminosities (if present) using our empirical correlations
constrained from a sample of radio quiet AGN (Eqs. 8-10). We
estimate the Eddington ratios, $L_{\rm bol}/L_{\rm Edd}$, from
estimated BH masses and bolometric luminosities, and find that all
CSS/GPS sources are high Eddington ratio systems
($10^{-2}$$<$$L_{\rm bol}/L_{\rm Edd}$$<$several, see Figure 4). The
mean Eddington ratios are $<\log L_{\rm bol}/L_{\rm Edd}>$=-0.55 and
-0.58 for CSS and GPS sources respectively, which are higher than
that of broad line radio galaxies $<\log L_{\rm bol}/L_{\rm
Edd}>$=-1.91 or radio loud QSOs ($<\log L_{\rm bol}/L_{\rm
Edd}>$=-1.11, e.g., Marchesini et al. 2004), but similar to those of
NLS1s ($<\log L_{\rm bol}/L_{\rm Edd}>\simeq$-0.8, e.g., Bian et al.
2008). Our results suggest that these young radio galaxies are also
in the early stage of the accretion activity and that the BHs are
still growing rapidly. Czerny et al. (2009) proposed that the
short-lived young radio galaxy may be due to the intermittent
activity caused by the radiation pressure instability in the
accretion disk, where the instability occurred when the accretion
rate (in the Eddington unit) exceeds a few percent. It should be
noted that most of the young radio galaxies in our sample are high
Eddington ratio sources (see Figure 4), which may have suffered the
radiation pressure instabilities. If this is true, the radiation
pressure instability will offer an interesting physical mechanism
for the possibly short-lived young radio galaxies (Wu 2009, in
preparation).

  Kawakatu et al. (2009) found that young radio loud AGN favor
an accretion disc without the big blue bump of the standard
accretion disc, based on photoionization modeling of various narrow
emission lines, and suggested that the accretion mode in these young
radio galaxies may be though advection dominated accretion flow
(ADAF, see Narayan \& McClintock for a recent review and references
therein). It is widely believed that optically thin ADAF exists only
for Eddington ratios less than a critical value $(L_{\rm bol}/L_{\rm
Edd})_{c}\simeq0.01$ (e.g., Narayan \& Yi 1995), and the accretion
mode should transit to an optically thick standard accretion disc
when the Eddington ratio larger than this critical value (e.g., Wu
\& Gu 2008 and references therein). However, our results show that
nearly all the young radio galaxies in our sample have Eddington
ratios larger than this critical value, and, therefore, it does not
support the ADAF scenario. The difference in the narrow line ratios
between young radio loud AGN and radio quiet Seyfert 2 galaxies in
Kawakatu et al. (2009) may be caused by other factors.

The width of the [O III] emission line has been used as a proxy for
stellar velocity dispersion to derive the BH masses in AGN, even
through the scatter is large (e.g., Nelson 2000; Boroson 2003;
Shields et al. 2003; Bonning et al. 2005). However, previous studies
of NLS1 galaxies, employing different samples and methods, led to
partially conflicting results in that the NLS1 galaxies are
systematically off the $M_{\rm bh}-\sigma_{*}$ relation if
$\sigma_{*}=\sigma_{\rm [O\ III]}$ is assumed (e.g., Bian \& Zhao
2004; Grupe \& Mathur 2004; Komosssa \& Xu 2007). The [O III] width
seems to be broader than gravitational velocities in these NLS1s. It
has been shown that CSS sources also tend to have rather broad [O
III] emission lines (e.g., Gelderman \& Whittle 1994; Nelson \&
Whittle 1996; Tadhunter et al. 2001; O'Dea et al. 2002; Holt et al.
2008). However, a direct comparison of [O III] width with the
central BH mass has not yet been made so far. We estimate the BH
masses for a sample of young radio galaxies (CSS/GPS sources) in
this work, and find that most of the sources lie systematically
\emph{off} the $M_{\rm bh}-\sigma_{*}$ relation, when using
$\sigma_{*}=\sigma_{\rm [O\ III]}$, as found for NLS1s (see Figure
6). A difference is that more than 50\% of BH masses in young radio
galaxies are larger than $10^{8}\msun$ and many of them have broad
line widths of $>2000\ \rm km\ s^{-1}$, while most BH masses of
NLS1s are less than $10^{8}\msun$ with broad line widths of $<2000\
\rm km\ s^{-1}$ (e.g., Grupe \& Mathur 2004). Therefore, the BH mass
is not the key parameter causing the deviation of [O III] emission
line width. We note that the extreme [O III] kinematic components,
as strong wings with blueshifts up to $\sim$2000 km/s, are found in
both radio quiet NLS1s (e.g., Aoki et al. 2005; Bian et al. 2005;
Komossa et al. 2008) and young radio galaxies (e.g., Tadhunter et
al. 2001; Holt et al. 2008), and these strong blue wings are often
not well included in the FWHM fittings. However, the sources with
extreme kinematics are still rare (e.g., Bian et al. 2005), which
will not change our main conclusion.

In the standard model, the narrow emission lines are excited through
photoionization by the UV continuum produced by the accretion disc.
The situation is more complicated in radio galaxies, since shocks
caused by the interaction of jet and interstellar medium may also be
important for the morphology, kinematics, and excitation of the
narrow emission lines. Inskip et al. (2002) investigated eight 6C
radio galaxies at redshift $z\sim1$ and concluded that combination
of accretion disc photoionization and shock photoionization provide
the best explanation of their spectra. Furthermore, by modeling the
ratios of different narrow emission lines with shock and AGN
photoionization models, Moy \& Rocca-Volmerange (2002) suggested
that the shock ionization may be dominant for radio sources with
intermediate radio sizes $D$ (2 kpc$<$$\it D$$<$150 kpc, e.g., CSS
sources), while AGN photoionization is dominant by the most compact
radio sources ($D$$<$2 kpc, e.g., GPS sources) and the extended
sources with $D$$>$150 kpc. We note that there are two GPS sources
(OQ 172 and 4C 12.50) with radio sizes of less one kpc, which also
show large deviations in their [O III] width (Figure 6). More
compact GPS sources are needed to further test this issue.

The jet-cloud interaction is expected to produce extreme kinematics
of narrow emission lines, and the jet-driven outflows provide a
convenient mechanism to explain the highly broadened line profile,
large velocity shift, and strong alignment between radio jet and
optical line in young radio galaxies (e.g., de Vries et al. 1997,
1999; Axon et al. 2000; Tadhunter et al. 2001; Holt et al. 2003,
2006, 2008; Labiano et al. 2005). We investigate the possible
relation between the deviation of [O III] line width $\Delta\sigma$
and the jet power $Q_{\rm jet}$ of young radio galaxies.  However,
there is no significant correlation between them based on our
sample. Therefore, the jet power should be not the only parameter
caused the extreme [O III] kinematics if the jet-driven scenario is
indeed important.

The Eddington ratio, which is proportional to the accretion rate per
unit BH mass, characterizes the extent to which radiation pressure
competes with gravity in the nucleus, and may also play an important
role in shaping the kinematics of the [O III] emission line apart
from the primary driver of gravitational potential of the bulge
(e.g., Greene \& Ho 2005b). After removing the blue wings, the core
component of $\sigma_{[\rm O\ III]}$ is still a good proxy of
stellar velocity dispersion $\sigma_{*}$ (e.g., Greene \& Ho 2005b;
Komossa \& Xu 2007). Blue wings are always found in NLS1s with high
Eddington ratios ($\sim 1$, e.g., Komossa et al. 2008), which may
result from the winds/outflows driven by the strong radiation
pressure in these high Eddington ratio sources. It is interesting to
note that most of young radio galaxies also have high Eddington
ratios, which are similar to those of NLS1s (Figure 3). We find that
the deviations of the [O III] emission line in these young radio
galaxies, $\Delta\sigma$, are also well correlated to the Eddington
ratios (see Figure 8). In particular, the young radio sources with
Eddington ratios $L_{\rm bol}/L_{\rm Edd}\sim1$ have the largest
deviations, which are also similar to those of radio quiet NLS1
galaxies (see Figure 8). The similarity in [O III] width deviations
between radio loud CSS/GPS sources and radio quiet AGN (QSOs and
NLS1s) suggests that the accretion activities may play a more
important role than the jet activities in shaping the kinematics of
[O III] narrow emission line, since the jet is absent or weak in
radio quiet AGN. The alignment of the narrow emission-line gas with
the radio jet is always used to as an evidence for the shock
scenario (e.g., de Vries et al. 1997; de Vries et al. 1999; Axon et
al. 2000; Privon et al. 2008). However, it should be also natural
for the alignment of radio jet and [O III] emission line if the
outflow is driven by the radiation pressure caused by accretion
activities.

\section{Summary}

Using a sample of 65 young radio galaxies (38 CSS + 27 GPS sources)
selected from the literature, we estimate BH masses, Eddington
ratios, and investigate the possible physical reasons for deviations
of width of [O III] narrow emission line. The main results can be
summarized as follows:

(1) The BH masses in young radio galaxies (CSS/GPS sources) range
from $10^{6.5}$-$10^{10.5}\msun$, which is the typical value in AGN
(see also Gu et al. 2009), and the mean value is $<$$\log M_{\rm
bh}$$>$$\simeq$8.3 which is less than that of radio loud QSOs or low
redshift radio galaxies ($<\log M_{\rm bh}>\simeq$9.0, e.g.,
Marchesini et al. 2004; Bettoni et al. 2003).

(2) Most of the CSS/GPS sources have relatively high Eddington
ratios, with mean value $<$$\log L_{\rm bol}/L_{\rm Edd}$$>$=-0.56,
which are similar to those of NLS1s. This result supports the idea
that the young radio galaxies are not only in the early stage of jet
activities but also in an early stage of accretion activities.

(3) The $M_{\rm bh}-\sigma_{\rm [O\ III]}$ relation in young radio
galaxies systematically deviates that defined by nearby inactive
galaxies, and reminiscent of NLS1s. We find that the deviation of
the [O III] width $\Delta\sigma$ is not related to the jet power or
accretion power, but is related to the Eddington ratio (Figures 6,
7, and 8). There is no evident difference between the radio loud
CSS/GPS sources and radio quiet QSOs/NLS1s in the
$\Delta\sigma$-$L_{\rm bol}/L_{\rm Edd}$ relation, which may suggest
that accretion activities may still play an important role in
shaping the kinematics of the [O III] emission line in these young
radio galaxies, since that radio jet is weak or absent in radio
quiet AGN.

\section*{ACKNOWLEDGMENTS}
   We thank the referee for constructive comments that helped to clarify the paper.
    We also thank Xinwu Cao, Andrew Humphrey and Minfeng Gu for their careful reading of the
   manuscript, valuable discussions and comments. This research has made use of the NASA/IPAC
    Extragalactic Database (NED) which is operated by the Jet Propulsion Laboratory,
    California Institute of Technology, under contract with the National Aeronautics and
    Space Administration. This work
is partly supported by the NSFC (grants number 10703009).

\end{document}